\documentclass{JHEP3}
\usepackage{cite}
\usepackage{graphicx}

\title{Magnetic monopoles in 4D: a perturbative calculation}
\author{Arsen Khvedelidze$^{ab}$ \\$^{a}$Department of Theoretical Physics, A.M.Razmadze Mathematical
              Institute, \\Tbilisi, GE-0193, Georgia\\E-mail: \email{akhvedilidze@plymouth.ac.uk}}
\author{Alex Kovner\\Physics Department, University of
Connecticut, 2152 Hillside Road, Storrs,\\ CT 06269-3046,
USA\\E-mail: \email{kovner@phys.uconn.edu}}
\author{David McMullan\\$^{b}$School of Mathematics and Statistics,
              University of Plymouth, Drake Circus,\\
             Plymouth PL4 8AA, UK\\E-mail:
             \email{dmcmullan@plymouth.ac.uk}
}

\abstract{We address the question of defining the second quantised
monopole creation operator in the 3+1 dimensional Georgi-Glashow
model, and calculating its expectation value in the confining phase.
Our calculation is performed directly in the continuum theory within
the framework of perturbation theory. We find that, although it is
possible to define the \lq\lq coherent state" operator $M(x)$ that
creates the Coulomb magnetic field, the dependence of this operator
on the Dirac string does not disappear even in the nonabelian
theory. This is due to the presence of the charged fields ($W^\pm$).
We also set up the calculation of  the expectation value of this
operator in the confining phase and show that it is not singular
along the Dirac string. We find that in the leading order of the
perturbation theory the VEV vanishes as a power of the volume of the
system. This is in accordance with our naive expectation. We expect
that nonperturbative effects will introduce an effective infrared
cutoff on the calculation making the VEV finite.}

\begin{document}

\section{Introduction}
Common lore has it that the condensation of magnetic monopoles is
responsible for confinement in nonabelian gauge
theories\cite{thooft}. Numerous lattice investigations of the
monopoles in pure Yang-Mills theories are available in the
literature \cite{lattice}. Nevertheless, there are many open
questions regarding the status of magnetic monopoles in confining
theories. The main problem with the monopole condensation scenario
is that no gauge invariant definition of the monopole operator can
be given in most theories of interest. Moreover, no gauge invariant
observable corresponding to magnetic charge exists at least in the
SU(2) pure Yang-Mills theory~\cite{us}. Thus serious doubts remain
regarding the significance of the monopoles for confinement in pure
gluodynamics.

On the other hand there undoubtedly exist theories where magnetic
monopoles can be given a proper gauge invariant meaning. A prime
example of such a theory is the Georgi-Glashow model in 3+1
dimensions. The model comprises of the SU(2) gauge field coupled to
the adjoint Higgs
\begin{equation}
L=-{1\over 4}F_{\mu\nu}^2+(D_\mu H)^2-V(H^2)\,.
\end{equation}
The theory is believed to have two phases as a function of the Higgs
potential $V(H^2)$. In the Higgs phase, where the VEV of the Higgs
field is nonvanishing, the magnetic monopoles exist in the spectrum
as heavy particles. Closer to the boundary of the confining phase
they become light, and it is natural to expect that they condense in
the confining phase of the theory. The magnetic charge (which, as
opposed to the pure Yang Mills theory, does exist here as a physical
observable) is then expected to be spontaneously broken in the
confining phase.

This behaviour has been all but established in the supersymmetric
cousin of the Georgi-Glashow model \cite{seibergwitten}. It is
however worth noting, that the confinement in the deformed N=2
supersymmetric theory is essentially abelian, and thus differs from
the confinement in QCD in many important aspects~\cite{yung}.
Confinement in the Georgi-Glashow model on the other hand is
expected to be fully nonabelian, and thus bear close similarity to
QCD.

Since the magnetic charge certainly exists as a physical observable
in the Hilbert space of the Georgi-Glashow model, one is tempted to
try to use this model as crutches on the way to the pure Yang-Mills
theory. More specifically, one would like to start with the
Georgi-Glashow model and trace the fate of the monopoles when the
mass of the Higgs field in the confining phase is taken to be very
large. In this limit the theory becomes the pure SU(2) Yang-Mills
theory. Of course, the perturbation theory, which is valid in the
Higgs phase, is not valid deep in the confining regime anymore.
Nevertheless, one can hope that the heavy Higgs field can be
integrated out perturbatively so that one can at least understand
what becomes of the monopole creation operator in the pure
Yang-Mills limit. Since we know that the magnetic charge disappears
in this limit, two options seem to be open for the monopole creation
operator. The first one is that it simply reduces to some
run-of-the-mill operator with the vacuum quantum numbers. This
operator could conceivably be a convenient order parameter for
confinement, even though it does not probe symmetry breaking any
more. The other possibility is that the condensate of the
Georgi-Glashow monopoles deep in the confining phase is so stiff,
that it costs the energy of the order of the Higgs mass to excite
it. In this case any excitation coupled to the condensate will be
very heavy, and in the limit the monopole creation operator would
simply reduce to a unit operator. In this case, it would become
useless for discussions of the properties of pure Yang-Mills theory.
In particular it could not be an order parameter for the deconfining
phase transition at finite temperature. The former option seems more
likely, since generically one expects light scalar excitations
(glueballs) to couple to the \lq\lq radial" part of the magnetic
monopole field, and thus to excite the condensate. Nevertheless the
second option is a logical possibility worth exploring/ruling out.

To study this question one has to start from the beginning, and the
beginning appears to be the construction of the monopole creation
operator in the continuum Georgi-Glashow model. Surprisingly,
although there exist a large number of lattice simulations in the
pure Yang-Mills theory, there have been to our knowledge no attempts
to explore this aspect of the Georgi-Glashow model. The continuum
literature is also sorely missing on this point. There have been
attempts to construct the monopole operator in the Georgi-Glashow
model \cite{marino}, but they appear to be incomplete. The aim of
the present paper is to construct a gauge invariant monopole
creation operator in the Georgi-Glashow model, and to set up the
perturbative calculation of its expectation value in the confining
phase. We construct a continuum gauge invariant operator which
creates the Coulombic magnetic field (without the Dirac string) of
the monopole and has the correct commutation relation with the
magnetic charge density:
\begin{eqnarray}
&&[M(x),B_k(y)]=\frac{i}{g}\,{(x-y)_k\over (x-y)^3}\, M(x)\,, \nonumber \\
&&[M(x),j_0^M(y)]={4\pi\over g}\, \delta^3(x-y)\, M(x)\,.
\end{eqnarray}
This operator is constructed by analogy with lower dimensional
theories as a \lq\lq coherent state" creation operator, that is an
exponential of an operator
 linear in electric fields.
An important property of the monopole operator is that it is not
rotationally invariant even though the magnetic field created by it
is spherically symmetric. This is directly related to the fact that
the Dirac string is present in the definition of the monopole
operator.  The direction of the string has to be specified and it
affects the commutators of $M$ with local gauge invariant
quantities, like the spacial components of electric current. We
stress that these commutators are not hopelessly divergent at the
position of the string, but merely depend on its orientation.

 We are also able to set up the perturbative calculation of the expectation value
 of $M$ in the confining phase. We find that, as in the calculation
of similar averages in lower dimensions \cite{kovner}, the relevant
path integral is dominated by a classical configuration. This
configuration interpolates in time between two states with magnetic
charge one half (in units of the magnetic charge of the
'tHooft-Polyakov monopole). At time plus infinity the magnetic
charge density is spread out homogeneously over space, while at
minus infinity the state contains a point-like 'tHooft-Polyakov
monopole and a balancing charge minus one half spread out
homogeneously in space. Viewed from a four dimensional perspective,
the configuration looks like radial magnetic current emanating from
the endpoint of a stem (world line of the 'tHooft-Polyakov
monopole). We therefore call this configuration \lq\lq a dandelion".
The physical amplitude described by this configuration corresponds
to the off diagonal matrix element between two components of the
vacuum wave function $\left<\Psi_1|M|\Psi_0\right>$, where
$\left|\Psi_1\right>$ is a state with magnetic charge one, and
$\left|\Psi_0\right>$ a state with magnetic charge zero. This is a
typical situation in theories with a spontaneous broken symmetry (in
the present case the symmetry generator being the magnetic charge).
A surprising property of the leading dandelion configuration is that
it is not spherically symmetric as one might naively expect. This is
the direct consequence of the lack of rotational invariance in the
definition of the monopole operator. The problem of string
dependence is not new and is well known in the framework of the
lattice gauge theory~\cite{latticestring}. A possible solution to it
has been suggested by Fr\"ohlich and Marchetti \cite{fm}. Although
the prescription of \cite{fm} directly applies to the calculation of
the correlation functions, and not to the operator itself, roughly
speaking it corresponds to the definition of $M$ as a weighted
average over operators which create two $Z_2$ strings rather than a
Coulomb magnetic field. This prescription goes outside the framework
of the coherent state operator, which we have adopted here and is
much more complicated. We therefore do not consider it in this
paper, and we believe that it will not alter qualitatively our
results.

Although the structure of the leading configuration suggests
spontaneous breaking of magnetic charge, we still find
$\left<M\right>=0$ even in the confining phase. The reason for this
is that the classical action of the dandelion is infrared divergent.
In particular, as a function of the infrared cutoff (the size of the
system $V$) we find $\left<M\right>\propto\exp\{-{\pi c\over g^2}\ln
V\}$, where $c$ is a pure number of order unity. This is not
unexpected for the simple reason that in the perturbation theory the
gluons are massless. In the presence of massless particles the
theory has no scale, and thus we expect a logarithmic divergence of
the classical action. The situation in the perturbation theory is
thus similar to the Kosterlitz-Thouless type phases in 1+1
dimensional theories \cite{kt}, where continuous symmetries are
\lq\lq almost" spontaneously broken and correlators of the order
parameter decay not exponentially, but rather as a power of the
distance. Beyond perturbation theory in the Georgi-Glashow model we
expect the situation to change. Nonperturbative effects will
introduce an infrared scale, the confinement radius, and we expect
this scale to cut off the infrared divergence of the classical
action and to lead to a finite VEV of $M$.

This paper is organised as follows. In Section 2 we discuss the
construction of the monopole operator. In Section 3 we set up the
calculation of its expectation value and discuss the properties of
the dandelion configuration. Section 4 contains some brief comments.

\section{The monopole creation operator}

The classical field configuration of the 'tHooft-Polyakov monopole is
\begin{equation}
H^a=h(r){r^a\over r}\,, \ \ \ A^a_i={f(r)\over
g}\epsilon_{aib}\,{r^b\over r^2}\,,\label{classical}
\end{equation}
where $h(0)=f(0)=0$,  $\lim_{r\rightarrow\infty}h(r)=v$,
$\lim_{r\rightarrow\infty}f(r)=1$ and $v$ is the vacuum expectation
value of the Higgs field. The distance scale on which the fields
approach their asymptotic values is given by their masses. Since we
are only interested in the infrared properties of the theory, we
will look for an operator which creates a point-like monopole and
thus from this point on fix $h(r)=v$, $f(r)=1$.

This configuration carries one unit of magnetic charge.
The gauge invariant magnetic charge in the Georgi-Glashow model can be defined in the
following way \cite{thooft}. First one defines a gauge invariant  field strength as
\begin{equation}\label{eq:tHooft}
\mathrm{F}_{\mu\nu}(x)= \displaystyle\frac{\,
H^a}{|{H}|}F^a_{\mu\nu}-
\displaystyle\frac{1}{g}\,\displaystyle\frac{1}{|H|^3}
\,\epsilon^{abc}\, H^a\,(D_\mu H)^b\,(D_\nu H)^c\,,
\end{equation}
in terms of the nonabelian field strength tensor
\[
F^a_{\mu\nu}=\partial_\mu\,A^a_\nu-\partial_\nu\,A^a_\mu +
g\,\epsilon^{abc}A^b_\mu\,A^c_{\nu}\,,
\]
and covariant derivatives of the  Higgs field $H^a$
\[
(D_\mu H)^a=\partial_\mu\,H^a + g\,\epsilon^{abc}A^b_\mu\,H^c\,.
\]
The  'tHooft tensor (\ref{eq:tHooft}) can also be written as
\begin{equation}
\mathrm{F}_{\mu\nu}=\partial_\mu\left(\hat H^a\,A^a_\nu\,
\right)-\partial_\nu\left(\hat
H^a\,A^a_\mu\right)-\displaystyle\frac{1}{g}\,\epsilon^{abc}\,\hat
H^a\partial_\mu\hat H^b\partial_\nu\hat H^c\,,
\end{equation}
with the unit vector field $\hat H^a={H^a/|H|}$. The magnetic
current is then defined as
\begin{equation}\label{magnc1}
k^M_\mu(x)=\frac{1}{2}\,\epsilon_{\mu\nu\lambda\sigma}\,\partial^\nu
\mathrm{F}^{\lambda\sigma}(x)\,.
\end{equation}
This definition is convenient since the magnetic charge density so
defined for all smooth configurations is equal to the topological
charge density of the Higgs field. The total magnetic charge is
then equal to the topological charge of $\hat H$.
\begin{equation}\label{qm}
Q_M=\frac{1}{4\pi}\,\int d^3x\, k^M_0\,(x) =
\frac{1}{8\pi\,g}\,\int \mathrm{d}^2
S_i\,\epsilon_{ij\,k}\epsilon^{abc} \hat H^a\partial_j\hat
H^b\partial_k\hat H^c\,,
\end{equation}
where the last integral is over the surface surrounding  the zeros
of the Higgs field. It should be noted that the definition
eq.~(\ref{magnc1}) is not unique. One can add to this current any
other conserved current with the same quantum numbers whose charge
vanishes on the 'tHooft-Polyakov monopole configuration. In
particular it is useful to consider
\begin{equation}
j^M_\mu=\epsilon_{\mu\nu\lambda\sigma}\partial^\nu \bigg(\hat
H^a(x)F^{\lambda\sigma\, a}(x)\bigg)\,.
\end{equation}
On the 'tHooft-Polyakov monopole configuration the charge
corresponding to this current is the same as $Q_M$ since the
covariant derivative of the Higgs field vanishes outside the
monopole core. On the other hand the density of this charge, unlike
that of eq.~(\ref{qm}), is not constrained to vanish outside the
points where $H=0$, and thus behaves like the majority of other
local operators. For this reason we prefer to use this alternative
definition of the magnetic current in this paper.

Our aim is to construct a quantum operator that creates a monopole
in the fully second quantised theory. This operator has to satisfy
the following commutation relations
\begin{equation}
[M(x),\,j^M_0(y)]=\frac{4\pi}{ g}\,\delta^3(x-y)\,M(x)\,.
\end{equation}
It is natural to constrain the operator $M$ further by requiring a somewhat more restrictive commutator
\begin{equation}\label{M}
M(x)B_k(y)M(x)^\dagger = B_k(y)+{i\over g}\,{(x-y)_k\over (x-y)^2}
\,,
\end{equation}
where $B_k=\frac{1}{2}\epsilon_{klm}\hat H^a F^a_{lm}$. This
latter commutator ensures that the operator $M$ creates a Coulomb
like magnetic field.

These commutation relations are reminiscent of the ones satisfied
by soliton creation operators in 2 and 3 dimensions \cite{kovner}.
There is however one crucial difference, and that is that the
monopole is a nonlocal object, and thus has a nonlocal commutation
relation with local gauge invariant fields, e.g. $B_i(x)$.

The most natural way of going about the  construction of  $M$ is to
try a coherent state like operator, that is an exponential of the
linear functional of electric fields. We thus start with the
following ansatz
\begin{equation}
M(x)=D(x)M_A(x)\,, \label{ans}
\end{equation}
where
\begin{equation}\label{oper}
M_A(x)=\exp\bigg(i\int \mathrm{d}^3y\, \lambda_i(x-y)\hat
H^a(y)E^a_i(y)\bigg)\,,
\end{equation}
with $\lambda_i$ the  classical vector potential of a point-like
Dirac monopole:
\begin{equation}
\lambda_i(x)={1\over g}\,\epsilon_{ij}\,{r^j_\perp\over
r^2_\perp}\,\left(\cos\theta-1\right)\,. \label{lambda}
\end{equation}
Here we have chosen the Dirac string to run in the direction of
the negative $x_3$ axis, and have defined  ${\bf
r}_\perp=(x_1,x_2)$. The operator $M_A$ creates a vector potential
of the Dirac monopole which in the isospace is oriented in the
direction of the Higgs field. In particular
\begin{equation}
M_A(x)B_i(y)M_A^\dagger(x) = B_i(y)+{i\over g}\left[{(x-y)_i\over
(x-y)^2}-4\pi\,
\hat{z}_i\,\delta^2(x_\perp-y_\perp)\theta(-(x_3-y_3))\right]\,.
\label{M1}
\end{equation}

Note that the operator $M_A$ is fully gauge invariant. It does
however create the magnetic field of the Dirac monopole including
the contribution of the Dirac string. As a result it does not create
a magnetic charge, since the magnetic flux due to the Coulomb
magnetic field is balanced by the flux of the Dirac string. The role
of the operator $D$ in eq.~(\ref{ans}) is to create the magnetic
field of the negative Dirac string alone, and thus cancel the
unwanted singular magnetic field in eq.~(\ref{M1}).

To construct the string operator $D$ we turn for inspiration to the
classical monopole solution eq.~(\ref{classical}). This solution is
written in the gauge where the vector potential is regular
everywhere except at the location of the monopole. One can however
transform this into unitary gauge, where the Higgs field is constant
everywhere in space. The transformation that achieves this is given
by
\begin{equation}
U(x)=\exp\{-i{\theta \over 2}\,\sigma\cdot\hat\phi\}\,,
\label{singu}
\end{equation}
where $\theta$ and $\phi$ are azimuthal and polar angles
respectively, $\sigma$ are Pauli matrices and the unit vector
\begin{equation}
\hat\phi^i=|x_\perp|\partial_i\phi=-\epsilon_{ij}{x_\perp^j\over
|x_\perp|}\,.
\end{equation}
The formal gauge transformation with
this gauge matrix gives
\begin{eqnarray}
&U^\dagger(x)A_i(x)U(x)&+\ {2i\over g}\,U^\dagger(x)\partial_iU(x)=\lambda_i(x)\sigma_3\,,\nonumber\\
&U^\dagger(x)\hat H(x)U(x)&=\ \sigma_3\,, \label{sing}
\end{eqnarray}
where we have defined $A_i\equiv A^a_i\sigma^a$ and $\hat
H\equiv\hat H^a\sigma^a$. The transformation eq.~(\ref{sing}) is not
a proper gauge transformation. The rotation matrix $U$ is singular
along the negative $x_3$ axis, where the direction of rotation
$\hat\phi$ is undefined. As a result the transformed configuration
is not equivalent to eq.~(\ref{classical}), as the magnetic field
corresponding to eq.~(\ref{sing}) has an extra Dirac string, and the
magnetic charge of eq.~(\ref{classical}) is cancelled by the string
contribution. The second quantised version of the transformation
eq.~(\ref{singu}) (or rather its inverse) is precisely what we need
to define to be able to get rid of the Dirac string contribution in
eq.~(\ref{ans}). It is however somewhat cumbersome to work directly
with the transformation eq.~(\ref{singu}) as it changes the gauge
variant variables everywhere in space. It is more convenient to
redefine it by a proper gauge transformation which achieves the same
result as eq.~(\ref{singu}) everywhere except infinitesimally close
to the Dirac string.

Define
\begin{equation}
\tilde U(x)=\exp\{-i{\tilde\theta \over 2}\,
\sigma\cdot\hat\phi\}\,, \label{regu}
\end{equation}
where the function $\tilde\theta$ is equal to the azimuthal angle
$\theta$ everywhere except inside an infinitely narrow tube around
the negative $x_3$ axis, e.g.
\begin{equation}
\tilde\theta=\theta\left[1-{1\over 2}(1-{\rm
sign}(x_3))f(x_\perp)\right],\ \ \ \ \ f(0)=1, \ \ \ \
f(x>\epsilon)=0\,. \label{reg}
\end{equation}
The action of this transformation on the configuration
eq.~(\ref{classical}) gives
\begin{eqnarray}
\label{nonsing} \tilde H&\equiv&\tilde U^\dagger \hat H\tilde
U=\sigma_3\cos(\theta-\tilde\theta)+\sigma \cdot\hat
r_\perp\sin(\theta-\tilde\theta)\,,\nonumber\\ \tilde
A_i&\equiv&\tilde U^\dagger(x)A_i(x)\tilde U(x)+{2i\over g}\,\tilde
U^\dagger(x)\partial_i\tilde U(x)\\\quad&=&-{1\over
g}\bigg(\partial_i\phi[\cos(\theta-\tilde\theta)-\cos\theta]\tilde{H}
-\partial_i(\tilde\theta-\theta)P_1-\partial_i\phi\sin(\theta-\tilde\theta)P_2
\bigg)\,, \nonumber
\end{eqnarray}
where
\begin{eqnarray}
&&P_1=\sigma\cdot\hat\phi\,, \qquad
P_2=\cos(\theta-\tilde\theta)\,\sigma\cdot\hat r_\perp
-\sin(\theta-\tilde\theta)\,\sigma_3\,.
\end{eqnarray}
In the formal limit $\tilde\theta=\theta$ the configuration
eq.~(\ref{nonsing}) becomes equivalent to eq.~(\ref{sing}), however
at any nonzero value of the regulator $\epsilon$ in eq.~(\ref{reg})
it is fully gauge equivalent to eq.~(\ref{classical}).

The transformation
\begin{equation}\label{v}
V(x)=\tilde U^\dagger(x)U(x)=\exp\{-i{\theta -\tilde\theta\over
2}\,\sigma\cdot\hat\phi\}
\end{equation}
is gauge equivalent to $U(x)$, but acts even on gauge variant fields
only in the infinitesimal neighbourhood of the negative $x_3$ axis.
To construct the  operator that creates the Dirac string we must
understand how to write down  the second quantised version of the
transformation eq.~(\ref{v}). The difficulty here is in the fact
that in writing down the matrix $V$ we have assumed explicitly that
the field $\hat H$ on which it is acting points in the third
direction in the isospace. For any other background, $V$ should be
additionally rotated by a regular, field dependent transformation.
It is however not very transparent how to deal with a transformation
whose parameters themselves depend on quantum fields. The
alternative is to take the limit $\epsilon\rightarrow 0$, in which
case the Higgs field does not change at all under the action of the
monopole operator. The vector potential however is affected by the
transformation $V$ even in this limit. Clearly, if $\hat H$ does not
change, the vector potential classically must acquire a non
singlevalued piece if the resulting magnetic field is to have a
nonvanishing divergence. It is not difficult to write down a
multivalued vector potential which gives a magnetic field of a Dirac
string.
\begin{equation}\label{mu}
\mu^a_3=\mu^a_\phi=0,\qquad\, \mu^a_r= {1\over g}\, \phi\,
\delta(r)\theta(-x_3)\, \hat H^a\,,
\end{equation}
where the radial delta function is one dimensional rather than two
dimensional to preserve the correct dimensionality of the vector
potential. A shift of $A$ by $\mu$ is affected by the following
second quantised operator
\begin{equation}
D(x)=\exp\bigg({i\over g}\,\int \mathrm{d}^3 y\, \phi(x-y)\,
\delta(|\mathbf{x}_\perp-\mathbf{y}_\perp|){(x_\perp-y_\perp)_i\over
|\mathbf{x}_\perp-\mathbf{y}_\perp|}\,\theta(-(x_3-y_3))\,\hat
H^a(y)E_i^a(y)\bigg)\,.\label{d}
\end{equation}
Even though this expression looks suspect, it is in fact a well
defined operator in a nonabelian theory. On the microscopic level
(ultraviolet cutoff scale) the electric field operator in a
nonabelian theory has a discrete spectrum which is quantised in
units of $g$. Thus the apparent ambiguity in the definition of
$\phi$ is not felt by any physical observable since it only leads to
ambiguity of an integer multiple of $2\pi$ in the phase of
eq.~(\ref{d}). The operator $D(x)$ is gauge invariant and is similar
in form to $M_A(x)$. In the following therefore we will not write it
down explicitly. We will continue denoting the vector potential of
the monopole by $\lambda_i$ but will understand it as the sum of
$\lambda_i$ of eq.~(\ref{lambda}) and $\mu_i$ of eq.~(\ref{mu}).

Hence, with this definition the monopole creation operator satisfies
the correct commutation relation with the abelian magnetic field,
eq.~(\ref{M}). This commutator is independent of the direction of
the Dirac string, even though the definition of $M$ uses explicitly
the Dirac string in a particular direction. However the dependence
on the string does not disappear from all commutators of $M$ with
gauge invariant operators. In particular consider the following
commutator
\begin{eqnarray}
M^\dagger_A(x)\left[ (B^a_i(y))^2-B_i^2(y)\right]M_A(x)&=&
\left[(B^a_i(y))^2-B_i^2(y)\right]+   2\epsilon_{ijk}
B^a_i(y)\lambda_j(y-x)(D_k\hat H(y))^a\nonumber\\&&\quad+
\left[\epsilon_{ijk}\lambda_j(y-x)(D_k\hat H(y))^a\right]^2\,.
\label{trans}
\end{eqnarray}
This explicitly depends on the vector potential $\lambda_i$.

We note that the commutator eq.~(\ref{trans}) has a direct analog in
the abelian theory. Consider an abelian gauge theory with a charged
scalar field $\phi$. The action of an abelian monopole creation
operator on the spatial components of electric current
$\phi^*D_i\phi$ is
\begin{equation}
M^\dagger(x)\left[\phi^*D_i\phi(y)-D_i\phi^*\phi(y)\right]
M(x)=\left[\phi^*D_i\phi(y)-
D_i\phi^*\phi(y)\right]+i\lambda_i(x-y)\phi^*(y)\phi(y)\,.
\end{equation}
Thus the action of the monopole creation operator produces a long range electric current.
Similarly the gauge coupled kinetic term of the scalar field transforms as
\begin{equation}
M^\dagger(x)D_i\phi^*D_i\phi(y)
M(x)=D_i\phi^*D_i\phi(y)-i\lambda_i(x-y)\left(
\phi^*D_i\phi(y)-D_i\phi^*\phi(y)\right)+\lambda_i^2\phi^*\phi\,.
\label{ab}
\end{equation}
It is this property of the magnetic monopole operator in theories
with charged matter fields that leads to dependence of the operator
on the choice of the string, as has been discussed extensively in
\cite{latticestring,fm}.

The commutator in eq.~(\ref{trans}) is very similar in physical
meaning. The operator $D_i\hat H$ is naturally identified with the
charged vector gauge boson fields $W^{\pm}$ (the two components of
the vector field perpendicular to the Higgs). The square of the
nonabelian magnetic field contains the minimal coupling of the
charged vector field $W^{\pm}$ to the abelian gauge potential. Thus
eq.~(\ref{trans}) is simply a transcription of eq.~(\ref{ab}) in
terms of spin one charged matter. The potentially disturbing
property of the commutator eq.~(\ref{trans}) is that it is much more
nonlocal than the commutator with the abelian magnetic field.
Whereas the abelian magnetic field created by the monopole operator
decreases as $1/(x-y)^2$, the nonabelian magnetic field decreases
only as $1/|x-y|$. As we shall see in the next section this
nonlocality has an important effect on the calculation of the
expectation value of the monopole creation operator in the confining
phase.

\section{Perturbative calculation of $\left< M\right>$}

In this section we set up the calculation of the expectation value
of the monopole creation operator in the confining phase. Although
we do not expect the perturbative result to be reliable, we do
expect to see substantial differences between the behaviour of the
expectation value in the confining and Higgs phases.

In the Higgs phase, since the monopoles are massive particles, the
correlation function of the monopole creation operator at large distance should behave as
\begin{equation}
\langle M^*(x)M(y)\rangle\propto \exp\big(- \mu|x-y|\, \big)\,,
\end{equation}
where $\mu$ is the monopole mass. Thus the expectation value in
the finite but large volume is
\begin{equation}
\langle M\rangle\propto\exp\big(-\mu L\big)\,.
\end{equation}

In the confining phase we expect the operator $M$ to have a finite expectation value.
However, since confinement is intrinsically nonperturbative, we still expect to
find a vanishing expectation value in the perturbative setup. However we do
expect the volume dependence at finite volume to be much milder.

Consider the path integral calculation of $\langle M\rangle$:
\begin{equation}
\langle M\rangle=\int D\hat HDA\exp\big(-S[A]\big)\,M[A]\ \sim
\int DA\exp\bigg(-{1\over 4}\int \mathrm{d}^4
x\,\big(F^a_{\mu\nu}-f^a_{\mu\nu}\big)^2\bigg)\,, \label{exp}
\end{equation}
with
\begin{equation}
f_{0i}^a=\lambda_i(x)\hat H^a\, \delta(x_4)\,,\qquad\, f_{ij}=0\,.
\label{cnumber}
\end{equation}
We are interested in calculating the expectation value deep in the
confining phase, where the Higgs field is very heavy. In the first
approximation we therefore neglect the Higgs contribution to the
action. The $c$-number field eq.~(\ref{cnumber}) is the shifted
Dirac vector potential, which multiplies the operator of the
electric field in the exponent eq.~(\ref{oper}). The extra
$c$-number contribution to the action $f^2$ arises when writing the
expectation value in the functional integral form as discussed in
detail for example in \cite{baruch}.

Path integrals of this type arise frequently in calculating
expectation value of operators which create topological excitations
\cite{kovner}. At weak coupling the path integral can be calculated
in the steepest descent approximation. This becomes obvious once one
realises that a simple rescaling of fields $A\rightarrow {1\over
g}A$, leads to the appearance of the factor $1/g^2$ in front of the
action in eq.~(\ref{exp}).

Our aim is therefore to find a classical configuration which
minimises the action in  eq.~(\ref{exp}). First we cast the path
integral expression into a more intuitively appealing form by
changing variables
\begin{equation}
A_\mu^a(x)\rightarrow A^a_\mu(x)-\lambda_\mu(\vec x)\theta(-x_4)\hat
H^a\,,
\end{equation}
with $\lambda_0=0$ and the spatial components of $\lambda_i$ given
by the sum of eq.~(\ref{lambda}) and eq.~(\ref{mu}).

Under this transformation
\begin{equation}
F^a_{\mu\nu} \rightarrow F_{\mu\nu}^a
-\bigg(\partial_{[\mu}\lambda_{\nu]}\hat H^a+\lambda_{[\nu}(D_{\mu
]}\hat H)^a \bigg) \theta(-x_4)-\lambda_{[\nu}\,\delta_{\mu ]4}\,
\hat H^a\,\delta(x_4)\,.
\end{equation}
 The path integral for the
calculation of VEV of the monopole operator becomes
\begin{equation}
\langle M\rangle=\int DA\exp\left\{-{1\over 4}
\left(F^a_{\mu\nu}-\bar f_{\mu\nu}\,\hat{H}^a
-\left(\lambda_{\nu}D_{\mu}\hat H-\lambda_{\mu}D_{\nu}\hat
H\right)^a\theta(-x_4)\right)^2\right\}\,, \label{exp1}
\end{equation}
where $\bar f_{\mu\nu}$ is the magnetic field of a point-like
'tHooft-Polyakov monopole at $x_4<0$,
\begin{equation}
\bar f_{ij}={1\over g}\,\epsilon_{ijk}\,{x_k\over
x^3}\,\theta(-x_4)\,.
\end{equation}

The general properties of the steepest descent configuration that
minimises the action in  eq.~(\ref{exp1}) can be understood using
the following argument. The insertion of the operator $M$ into the
path integral annihilates a magnetic monopole at time $t=0$. Thus
one expects that the leading configuration has a smooth magnetic
charge density which at negative times integrates to a different
value than at positive times, the difference being the magnetic
charge of the monopole annihilated by $M$. One might expect that the
action is minimised by a spherically symmetric configuration. The
natural candidate for such a magnetic charge/current density would
be
\begin{equation}
j_\mu(x)={1\over g}\,{x_\mu\over |x|^4}\,, \label{spher}
\end{equation}
since it has a divergence at the point in space-time where the
monopole  is created, and thus is naturally associated with a
space time event of annihilation of a monopole. The overall
coefficient is determined by the requirement that the 3D flux
emanating from this point is the same as that created by $M$. Such
a current cannot however be constructed from the field strength at
all, since by definition any magnetic current that can be written
in terms of the fundamental fields of the theory is
divergeneseless. Thus one needs to balance the total flux that
emanates from the point $x=0$. The simplest and the most natural
way to do that is to have the balancing flux going  in an
infinitely thin flux tube:
\begin{equation}\label{current}
j^M_\mu={1\over g}\left[{x_\mu\over |x|^4} -4\pi\,\delta_{\mu 4}\,
\delta^3(\vec x)\, \theta(-x_4)\right]\,.
\end{equation}
The thin flux tube term in eq.~(\ref{current})  has a natural
interpretation of the world line of the monopole  annihilated by the
operator $M$. This structure is very similar to the one arising in
the calculation of the VEV of the magnetic vortex operator in 2+1
dimensions \cite{kovner} and is generic for a broken symmetry phase.
Pictorially it represents the flux of the spontaneously broken
charge emanating from a single space-time point in a spherically
symmetric way, like the head of a dandelion, while the necessary
\lq\lq nutrients" (incoming flux) are supplied by a thin stem
extending infinitely into the past. We will thus refer to a
configuration of this generic type  as a \lq\lq dandelion".

We will find however that a spherically symmetric  dandelion does
not minimise the classical action. Instead the action is lower for
an axially symmetric dandelion, the spherical symmetry being broken
due to the sensitivity to the direction of the Dirac string. In the
rest of this section we will construct the vector potential for an
axially symmetric dandelion and consider the minimisation of the
action. Although we cannot prove that the configurations of the type
we consider lead to an absolute minimum of the action, we believe
that parametrically (and physically) our results are correct.

We start with the following expression for an  axially symmetric
vector potential and the Higgs field  in four Euclidean
dimensions:
\begin{equation}\label{ansatz}
\hat H^a={x^a\over r}\,, \qquad   A^a_i ={f(z, x)\over
g}\,\epsilon_{aib}{x^b\over r^2}\,,
\end{equation}
where
\begin{equation}
r^2=x_1^2+x_2^2+x_3^2,\ \ \ \ z={x_4\over r}\,, \qquad\,
x=\cos\theta\,.
\end{equation}

Calculating the components of the field strength we find
\begin{eqnarray}\label{fields}
F^a_{ij}&=&F_{ij}\hat H^a +{1\over g}{x_k\over r^3}
\left[\epsilon^{ajk}\left(-{x_i\over r}\, z{\partial
f\over\partial z}+
(\delta_{i3}-\frac{x_3 x_i}{r})\,
{\partial f\over\partial x}\right) - \ (i\, \leftrightarrow\,  j)\ \right]\,,\\
F^a_{4i}&=&\partial_4 A_i^a={1\over g}\epsilon_{aib}{x^b\over
r^2}\partial_4 f(z,x)= {1\over g}\, \epsilon_{aik}{x_k\over r^3}\,
{\partial f\over\partial z} \,,
\end{eqnarray}
where
\begin{equation}\label{eq:abpro}
    F_{ij} = -{1\over g}{\epsilon_{ijk}x_k\over r^3}\,(2f-f^2)\,.\\
\end{equation}
The covariant derivative of the Higgs field is given by
\begin{equation}
(D_i H)^a= {1\over r}\,(\delta^{ai}-{x_i x_a\over r^2})(1-f)\,.
\end{equation}
The magnetic current for this configuration is
\begin{equation}
j^M_\mu={1\over g}\,\left[{x_\mu\over r^4}(1-f){\partial f\over
\partial z}- 4\pi \delta_{\mu 4}\, \delta^3(\vec
x)\theta(-x_4)\right]\,.
\end{equation}
This looks like a radial dandelion but with the density  of the
current not spherically symmetric - a dandelion in the wind.

With these expressions we find  that the classical action  in
eq.~(\ref{exp1}) is given by the following integral
\begin{equation}
S_{\rm{dandelion}}=-{\pi\over g^2}\int_0^\infty {dr\over
r}\left[S_1+S_2+S_3+S_4+S_5\right]\,,
\end{equation} with
\begin{eqnarray}
S_1&=&\int_{-\infty}^{\infty} dz\int_{-1}^{1}dx\left[2f-f^2-\theta(-z)\right]^2\,,\\
S_2&=&\int_{-\infty}^{\infty} dz\int_{-1}^{1}dx\left({\partial f\over\partial z}
\right)^2\,,\nonumber\\
S_3&=&\int_{-\infty}^{\infty} dz\int_{-1}^{1}dx\,2z^2\left({\partial f\over\partial z}
\right)^2\,,\nonumber\\
S_4&=&\int_{-\infty}^{\infty} dz\int_{-1}^{1}dx(1-x^2)\left({\partial f\over\partial x}
\right)^2\,,\nonumber\\
S_5&=&\int_{-\infty}^{\infty}
dz\int_{-1}^{1}dx\,\theta(-z)(1-x)(1-f) \left[{1\over
1+x}(1-f)-2{\partial f\over\partial x}\right]\,.\nonumber
\end{eqnarray}
The term $S_1$ comes from the square of the abelian  component of
chromomagnetic field, the term $S_2$ from the chromoelectric field,
while $S_3+S_4+S_5$ arise due to the square of the nonabelian part
of the chromomagnetic field, the square of the term in
eq.~(\ref{exp1}) which involves the covariant derivative of the
Higgs field, and the interference term between the two. Defining
\begin{equation}
K=1-f\,,
\end{equation}
we can rewrite
\begin{eqnarray}
S_{\rm{dandelion}}&=&-{\pi\over g^2}\ln (\Lambda
L)\int_{-\infty}^{\infty}
dz\int_{-1}^{1}dx\left\{[1-K^2-\theta(-z)]^2+ \left({\partial
K\over\partial z}\right)^2(1+2z^2)\right.\nonumber
\\
&&\quad+\left.(1-x^2)\left({\partial K\over\partial x}\right)^2
+\theta(-z){1-x\over 1+x}\left[K^2+(1+x){\partial K^2\over\partial
x}\right]\right\}\,,
\end{eqnarray}
with $\Lambda$  an ultraviolet cutoff, and $L$  the linear size of
the system.

The ultraviolet divergent has to do with the  fact that our
monopole has a zero core size, and is therefore not a problem. The
infrared divergence appears since our ansatz for the dandelion
vector potential is dilatationally invariant. As we will argue
shortly, this is unavoidable and indicates that in perturbation
theory the expectation value vanishes if the volume of space time
is infinite. We believe this is an artifact of the perturbative
calculation, and nonperturbative confinement effects must provide
an effective infrared cutoff on this calculation, regulating this
divergence.

The most important question is of course whether the remaining $x$
and $z$  integrals are finite. It is easy to see that this is the
case for any smooth function which satisfies the following
boundary conditions
\begin{eqnarray}
&&K(z,-1)=0\,,  \qquad z<0\,, \nonumber\\
&&|K(z\rightarrow\infty, x)-1|<{1\over z}\,,\\
&&|K(z\rightarrow-\infty, x)|<{1\over z}\,.\nonumber
\end{eqnarray}
We take as an ansatz for $K$ functions of the form
\begin{equation}
K(z)=\rho(z)(1+x)^{1-\rho(z)}\,,
\end{equation}
where $\rho(z)$ is a smooth approximation to the step function. As
an example consider the one parameter class of functions
\begin{equation}\label{eq:trial}
\rho(z,\gamma):=\frac{1}{2}+ \frac{2}{\pi}\arctan\bigg(\frac{\gamma
z}{\sqrt{1+\gamma^2 z^2}}\bigg)\,.
\end{equation}
From these we can construct a two-parameter class of functions
$K(z,x,\alpha,\beta)$ where
\begin{equation}
    K(z,x,\alpha,\beta) =\rho(z,\alpha)(1+x)^{1-\rho(z,\beta)}\,.
\end{equation}
Writing
$S_{\mathrm{dandelion}}(\alpha,\beta)=-\displaystyle\frac{\pi}{g^2}\,\ln(\Lambda
L)c(\alpha,\beta)$ and calculating the integrals numerically, we
find as shown  in Figure~1,  a minimum value for the action at
$\alpha\approx1.35$ and $\beta\approx2.65$  corresponding to
$c(\alpha,\beta)\approx2.17$. Several other smooth approximations to
the step function have been investigated and all lead to a higher
action.
 \FIGURE[ht]{
\includegraphics{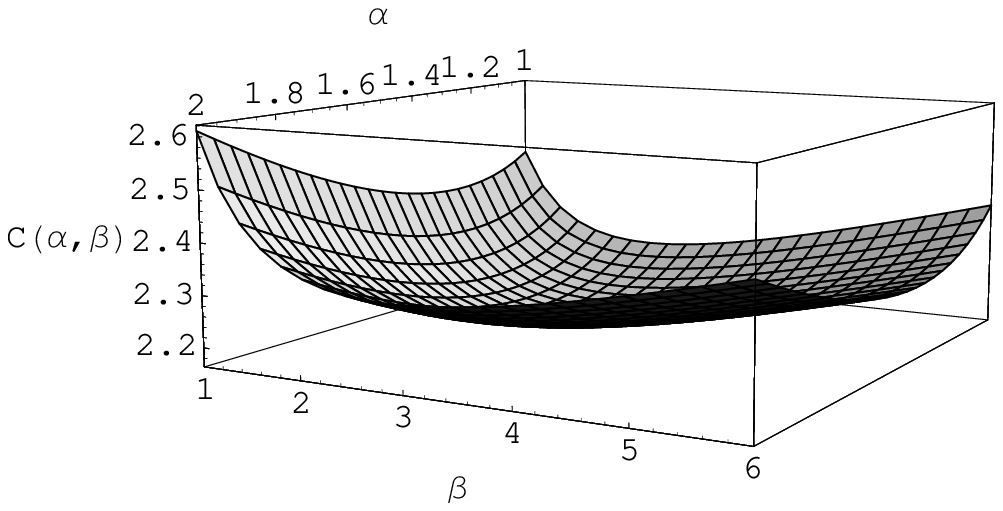}
\caption{Plot of the two parameter variation $c(\alpha,\beta)$ for
the dandelion action.}}

A striking property of these boundary conditions is that a
spherically symmetric ansatz, corresponding to the function $K$
independent on the angle $x$ has an action which diverges worse
than logarithmically. In particular if we take
\begin{equation}
K=\sqrt {1-{1\over \pi}\left({\rm arctan}{1\over z}-{z\over
1+z^2}\right)}\,,
\end{equation}
which gives the spherically symmetric magnetic current
eq.~(\ref{spher}), the $z,x$ integral diverges as the second power
of the infrared logarithm. The spherically symmetric configuration
thus gives a negligible contribution to the path integral for the
expectation value of $M$.

Physically the meaning of the dandelion configuration is quite
clear. As $x_4\rightarrow-\infty$, the magnetic field
eq.~(\ref{fields}) is the field of the point-like 'tHooft-Polyakov
monopole. At negative but non-infinite times an extra magnetic
charge density appears at distance scales of order $x_4$. This extra
charge density is not spherically symmetric but rather has only
axial symmetry and is pushed away from the Dirac string. Since our
ansatz eq.~(\ref{ansatz}) is dilatationally invariant, we expect
that for any field of this type the extra magnetic charged density
is small within the cone of the opening angle $x_4/r$ around the
string.  As $x_4\rightarrow\infty$, the field vanishes. Thus from
this perspective it looks like a configuration that interpolates
between a monopole in distant past and a vacuum in distant future.
Note that even so, the conservation of the magnetic current cannot
be violated by the dandelion. To see this explicitly we have to
analyse the field profile at fixed $x_4$. As long as $r<|x_4|$, the
field is what we have just described - a monopole at negative time,
and vacuum at positive time. However at $r>|x_4|$, the field is
given simply by the value of the function $K$ at $z=0$. As long as
$K$ is a continuous function of $z$, this means that the magnetic
flux at spatial infinity is time independent. The balancing flux is
precisely due to the extra magnetic charged density which lives at
distances of order $r\propto x_4$. Thus the dandelion configuration
describes at negative times a point-like   'tHooft-Polyakov
monopole, which at radius $r\sim|x_4|$ is surrounded by an axially
symmetric \lq\lq shell" of negative magnetic charge density. As
$|x_4|\rightarrow 0$ the radius of the shell shrinks to zero. At
positive times the configuration has no core but instead all the
magnetic charge density is concentrated in a shell whose size grows
as $x_4\rightarrow\infty$.

Note that the logarithmic  infrared divergence is not an artifact
of our dilatationally invariant ansatz. It is easy to see that one
cannot do better than that. Let us relax the assumption that $K$
depends only on the ration $x_4/r$. A little algebra then leads to
the following expression for the action of the dandelion
configuration
\begin{eqnarray}\label{nondil}
S_{\rm{dandelion}}&=&-{\pi\over g^2}\int dr \int_{0}^{\infty}
dx_4\int_{-1}^{1}dx\left[{1\over r^2}(1-K^2)^2
+\left({\partial K\over\partial x_4}\right)^2+2\left({\partial K\over\partial r}
\right)^2\right.\nonumber\\&&\hspace*{7cm}+\left.{1\over r^2}(1-x^2)\left({\partial K\over\partial x}\right)^2\right]\\
&-&{\pi\over g^2}\int dr\int_{-\infty}^{0} dx_4\int_{-1}^{1}
dx\left[{1\over r^2}K^4 +\left({\partial K\over\partial
x_4}\right)^2+2\left({\partial K\over\partial
r}\right)^2\right.\nonumber\\&&\hspace*{7cm}+\left.{1\over
r^2}{1-x\over 1+x}\left(K+(1+x){\partial K\over\partial
x}\right)^2\right]\,.\nonumber
\end{eqnarray}
Let us follow the change of K with time $x_4<0$ at fixed value of
$r$.  The qualitative behaviour is clear from our discussion above.
At large negative times $K=0$, while at times close to zero, $K$ is
a fixed function $\kappa(x)$ which does not vanish at large $r$.
Let's call the time interval during which this change in behaviour
occurs $\Delta T(r)$. We can estimate the magnitude of various terms
in eq.~(\ref{nondil}). In particular
\begin{eqnarray}
&&\int_{-\infty}^{0} dx_4{1\over r^2} K^4\propto {\Delta T(r)\over r^2}\kappa^4\,,\\
&&\int_{-\infty}^{0} dx_4\left({\partial K\over\partial
x_4}\right)^2 \propto {1\over \Delta T(r)}\kappa^2\,.\nonumber
\end{eqnarray}
It is obvious from these two expressions  that
$T(r)\rightarrow_{r\rightarrow\infty}Ar$, or else the subsequent
integral over $r$ diverges stronger than logarithmically.

Thus we conclude that in the leading order  in perturbation theory
the expectation value of the monopole operator is
\begin{equation}\label{expv}
\left<M\right>=\exp\bigg(-{\pi c \over g^2}\ln (\Lambda L)\bigg)
\end{equation}
with constant $ c\approx 2.17$.

\section{Discussion}

In this paper we have shown how to set up the perturbative
calculation of the expectation value of the monopole creation
operator in the confining phase. Perturbatively we find that the VEV
vanishes in the infinite volume. However our result eq.~(\ref{expv})
is an intrinsically perturbative one, and as such we certainly do
not expect it to stand beyond perturbation theory. Nonperturbative
contributions will provide an infrared cutoff. It is thus very
likely that the $\left<M\right>\ne 0$ in the confining phase.

A perturbative calculation of VEV is of course not reliable  since
it is dominated by the infrared physics. A quantity which is more
amenable to such a calculation is short distance behaviour of the
monopole-antimonopole correlation function~\cite{aad}. Although we
have not calculated this quantity, the perturbative cutoff
dependence of $\left<M\right>$ suggests that the correlator behaves
as
\begin{equation}
\left<M(x)M^\dag(y)\right>\, \propto \,  \displaystyle | x- y
|^{-{\pi c\over g^2}}\,.
\end{equation}
This indicates that even in the pure  Yang-Mills limit the monopole
operator does not freeze, even though its phase must decouple from
the finite mass spectrum. Instead $M$ couples to finite mass
excitations via its modulus.

It is worth noting that the present approach  can be easily extended
to finite temperature. It is interesting to see whether at high
temperature the VEV of $M$ vanishes in perturbation theory.

\acknowledgments
 We are grateful to Gerald Dunne and Vishesh Khemani for useful discussions.

\end{document}